\documentclass[12pt]{emulateapj}

\usepackage[dvips]{color}

\shorttitle{3C~84 jet}
\shortauthors{Kino et al.}

\begin{document}

\title{Evidence of jet-clump interaction: a flip of the radio jet head of 3C~84}
\author{
M.~Kino\altaffilmark{1,2},
K.~Wajima\altaffilmark{3},
N.~Kawakatu\altaffilmark{4},
H.~Nagai\altaffilmark{2},
M.~Orienti\altaffilmark{5},
G.~Giovannini\altaffilmark{5,6},
K.~Hada\altaffilmark{2},
K.~Niinuma\altaffilmark{7},
M.~Giroletti\altaffilmark{5}}

\altaffiltext{1}{Kogakuin University of Technology \& Engineering, 
Academic Support Center,
2665-1 Nakano, Hachioji, Tokyo 192-0015, Japan}
\altaffiltext{2}{National Astronomical Observatory of Japan,
                 2-21-1 Osawa, Mitaka, Tokyo, 181-8588, Japan}
                 \email{motoki.kino@nao.ac.jp}
\altaffiltext{3}{Korea Astronomy and Space Science Institute, 
776 Daedeokdae-ro, Yuseong-gu, Daejeon 34055, Republic of Korea} 
\altaffiltext{4}{National Institute of Technology, Kure College, 2-2-11, 
Agaminami, Kure, Hiroshima, 737-8506, Japan}
\altaffiltext{5}{
INAF - Istituto di Radioastronomia, 
via Gobetti 101, 40129 Bologna, Italy}
\altaffiltext{6}{
Dipartimento di Astronomia, Universita' di Bologna, via Gobetti 103/2, I-40129, Bologna, Italy}
\altaffiltext{7}{
Graduate School of Sciences and Technology for Innovation, Yamaguchi University, 1677-1 Yoshida, Yamaguchi, Yamaguchi 753-8512, Japan }



\begin{abstract}

Radio jets in active galaxies have been expected to 
interact with circumnuclear environments in their early phase evolutions.
By performing the multi-epoch monitoring observation
 with the KVN and VERA Array (KaVA)  at 43~GHz,
we investigate the kinematics of the notable newborn bright  component C3 
 located at the tip of the recurrent jet of 3C~84.
During 2015 August-September, we discover 
the  flip of  C3 and the amount of the  flip is about 0.4~milli-arcsecond  in angular scale, 
which corresponds to 0.14 parsec in physical scale.
After the flip of C3, it wobbled at the same location for a few months
and then it restarted to propagate towards the southern direction.
The flux density of C3 coherently showed the monotonic increase during the observation period.
The flip is in good agreement with hydrodynamical simulations of jets in clumpy ambient medium.
We estimate the number density of the putative clump based on the momentum balance
between the jet thrust and the ram pressure from the clump  and it is  about $10^{3-5}~{\rm cm^{-3}}$.
We briefly discuss possible origins of the clump.

\end{abstract}

\keywords{
galaxies: active 
radio continuum: galaxies
gamma rays: galaxies 
galaxies: individual (3C~84, NGC~1275)}

\section{Introduction}

Active galactic nuclei (AGNs) are powered by surrounding gas accretion
onto supermassive black holes (BHs) at the center of each galaxy.
Since the  circumnuclear material plays important roles of 
a gas reservoir for the accretion flow, 
its nature, such as the structure, the size, and the mass,
has been intensively investigated (e.g., Ramos Almeida and Ricci 2017 for review).
While AGNs themselves radiate across the entire electromagnetic spectrum,
from the radio and up to $\gamma$-rays, 
circumnuclear material do not emit significant electromagnetic signal.
Hence, circumnuclear matter distributions in particular at the central 
parsec scale regions in AGNs are highly uncertain.
Observationally, there are some indications of inhomogeneous matter distribution.
VLBI observations of young  radio sources 
reveal that a large fraction of young radio lobes with a two-sided structure
show the asymmetry in arm-length ratio and their flux density from a pair of the lobes 
and the pair of the lobes satisfies a brighter-when-farther behavior
 (Dallacasa and Orienti 2013; Orienti and Dallacasa 2014; Orienti 2016).
This asymmetry can be naturally explained by the interaction between the jet 
and an inhomogeneous environment.
At X-ray energy band,
time variations in absorbing column density on timescales from
months to years have been known, which indicate
highly non-uniform circumnuclear material such as clumps 
on a sub parsec scale (Ives et al. 1976; Malizia et al. 1997; Risaliti et al. 2002).
Theoretically,
jets are thought to have a strong impact on the interstellar medium of a host radio galaxy.
The deposition of jet mechanical energy may affect  formation of stars in the host galaxy and 
 the accretion of matter down the gravitational potential onto the central BH,
 which is so called AGN feedback 
 (e.g., Silk and Rees 1998; 
 Bicknell et al. 2000; 
 King 2003; 
 Croton et al. 2006; 
 Fabian 2012). 
In the context of AGN feedback, Wagner and Bicknell (2011) ran numerical simulations of
AGN jets interacting with a non-uniform medium containing dense clumps, 
focusing in the effects of the AGN jet on the cold dense clumps in which stars were formed.
Intriguingly, Wagner and Bicknell (2011) indicate flips of 
the jet head caused by the  interaction with the clumps,
although such a phenomenon has not yet been observed so far.
Similar behavior have been also reported in several other studies (e.g., Fragile et al. 2017).
However, no direct evidence for such dramatic impact is yet reported so far.

NGC~1275  is a notable nearby giant elliptical galaxy 
at the core of the Perseus cluster, with an optically luminous nucleus, 
currently classified as a Seyfert 1.5/LINER (Sosa-Brito et al. 2001). 
Interestingly,  Nagai et al. (2010) found the emergence of newborn
 bright component so-called C3 during
  multi-epoch monitoring of 3C~84 using Very Long Baseline Interferometer (VLBI).
The component C3 showed a proper motion towards the southern direction
with the apparent velocity of $0.2-0.3~c$ (Nagai et al. 2010; Suzuki et al. 2012; Hiura et al. 2018).
Since various observations of 3C~84 indicate the gas-rich environment such as
molecular gas (Krabbe et al. 2000; Salome et a. 2006; Lim et al 2008),
warm [H$_{2}$] gas (Wilman et al. 2005; Scharwachter et al. 2013)
and dense ionized plasma surrounding the 3C~84 jet (O'Dea 1984; Walker et al. 2000), 
3C~84 is regarded as a quite unique laboratory to explore  interactions 
between the jet and the circumnuclear  environment on parsec scale.
In this work, we will report the KaVA monitoring observation of 3C~84 
during 2015-2016 , in which we indeed discover a theoretically 
predicted a flip of the jet head  accompanying  
the clear increase of the C3 flux density.


In this work, we define the radio spectral index $\alpha_{R}$
as $S_{\nu}\propto \nu^{-\alpha_{R}}$.
The cosmological parameters used here are as follows;
$H_{0} = 71~{\rm km/s/Mpc}$, 
$\Omega_{\lambda} = 0.73$ and 
$\Omega_{m} = 0.27$
(e.g., Komatsu et al. 2011).
The redshift of 3C~84 ($z=0.018$) is located 
at the distance  $75$~Mpc and 
it corresponds to 0.35~pc mas$^{-1}$.
The mass of the black hole in NGC~1275 is estimated to
be around $M_{\bullet} \approx 8 \times 10^{8}M_{\odot}$ 
(Schawachter et al. 2013) and we adopt this value.
Corresponding Eddington luminosity is
$L_{\rm Edd}\approx 1 \times 10^{47}~{\rm  erg~s^{-1}}$.
Normalized physical quantity is denoted 
as $Q=10^{x}Q_{x}$, otherwise stated.

\section{Observation and data reduction}

From 2015 August to 2016 June,
we conducted  full-track 12~epochs monitoring of 3C~84 at 43~GHz
with  KaVA, the combination
of the Korean VLBI Network (KVN) and
the Japanese VLBI Exploration of Radio Astrometry
(VERA) radio arrays.
The joint array, KaVA, has a total of seven antennas
at frequencies of 22 and 43~GHz, resulting in angular
resolutions of about 1.2 and 0.6~mas, respectively. 
Basic imaging capabilities of KaVA has been summarized
in detail by Niinuma et al. (2014) and Hada et al. (2017).
Left-hand circular polarization was received and sampled with a 2 bit quantization. 
All the data were recorded at 1~Gbps (256~MHz bandwidth,
16~MHz, 16 sub-bands) and correlated by the Daejeon correlator.
As a technical note, we note that 
in the flux calibration of KaVA one
 visibility amplitudes of KaVA
 correlated by the Daejeon hardware correlator (Lee et al. 2015a)
 should be corrected by multiplying the factor of 1.35 
(Lee et al. 2015b).

In Figure~\ref{fig:visibility}, 
we show the visibility amplitude and phase obtained in 
one of the KaVA monitoring epochs with a typical data quality.
On-source time for 3C~84 was about 6-7~hours
and that is sufficiently long for obtaining a good uv-coverage.
The typical size of the original beam of KaVA at 43~GHz is 0.6-0.7~mas and 
we restored all of the images with the beam size of 0.75~mas. 
In Table 1, we show the observation summary dealt in this work.
As seen in the second column of the Table~1, some stations 
were lost (station(s) followed after the minus sign mark in the second column) 
in some of the KaVA monitoring  observations due to system troubles in some of the epochs.
Three epochs of the  images obtained by the Very Long Baseline Array (VLBA)
in the framework of the Boston University (BU) Blazar monitoring program
(https://www.bu.edu/blazars/research.html) 
are also included for consistency check purpose.
We followed the standard procedures for initial phase and
amplitude calibration using the National Radio Astronomy 
Observatories (NRAO) Astronomical Imaging Processing 
System (AIPS) (Greisen 2003).
{\it A priori} amplitude calibration was applied using the measured 
system noise temperature and the
elevation-gain curve of each antenna. We calibrated the 
bandpass characteristics of phase and amplitude at each station
using the auto-correlation data. 
Following the amplitude calibration,
fringe-fitting was performed to calibrate the visibility phases,
and finally the data were averaged over each 
intermediate frequency band.
The  imaging and self-calibration were performed in the Difmap 
software (Shepherd 1998) in the usual manner.

\section{Observational results}

\subsection{Position of C3 component}

Let us estimate the positional accuracy of C3 relative to C1.
\footnote
{Since we did not conduct a phase-referencing mode observation but 
a normal imaging one, the position of C3 is  defined as a relative position to C1.}
The peak positions  of the C1 and C3 components
are determined based on the standard model-fitting procedure in Difmap.
First, we estimate a systematic error in position due to 
the positional scattering of the peak position of  C1 
from the origin-point in the difmap  during the period 
when the peak flux density of C1 is brighter than that of C3
(first half of our observation period).
From the data, we obtained the positional scattering of C1
about $0.02-0.07$~mas.
Second, we should  estimate the systematic error.
A deconvolution error is known as a major systematic error  and it is
possibly caused by some other component such as  the C2 component (e.g.., Nagai et al. 2010). 
In our observation, however, the peak flux density of  C3 ($\gtrsim 5$~Jy) is more than
the factor of $\sim 3$ brighter 
than that of C2  ($0.9-1.8$~Jy) and the separation angle of C2 and C3 ($\sim 2$~mas)
is larger than the beam size of KaVA at 43~GHz ($\sim 0.6-0.7$~mas).
Therefore, it is clear that the deconvolution error by C2 is negligibly small.
Third, we add to note that a nominal estimate of the position accuracy, 
i.e.,  the KaVA beam size divided by  the dynamic range (a few $10^{3}$)
is even much smaller than the systematic error by the peak position scattering of C1.
Based on these error estimations, 
we conservatively set the positional accuracy of C3 relative to C1 as $0.07$~mas.
The  error bars in the position in Figure~\ref{fig:multi} correspond to the  $0.07$~mas.
Since both of C1 and C3 are so bright that 
one can generally obtain a quite high positional accuracy of C3 relative C1,
which is clearly reflected in the high values of the dynamic ranges shown in the Table~1.

For consistency check purpose, we have included three epochs of 
VLBA archival data from the Boston University (BU) Blazar monitoring program.
The peak positions of  C3 with VLBA (see Appendix) 
are also shown in Fig.~\ref{fig:position} with the triangle mark 
and these locations are consistent with those measured with KaVA.

Generally, one should be aware that C3 has substructures in it 
(e.g., Nagai et al. 2014; Giovannini et al. 2018).
When we discuss these substructures inside the C3 component,
then a higher spatial resolution is required since substructures 
can not be resolved with the KaVA beam. 
In the present work, we focus on the kinematics of the spatially-averaged 
positions of  C3 and thus the KaVA beam size is enough for this investigation.
We also confirm good agreements of spatially-averaged positions in C3 
between KaVA and VLBA images shown in Appendix. 

In terms of structures around C3, we should also note a newly generated 
temporal structure shortly extending towards the West of C3 in the subsequent 
three epochs (2015 Dec 6th, 2016 Jan 5th, and Jan 20).
We will briefly discuss a possible origin of this substructure  in \S 5.2.

\subsection{Flip of the C3 component}\label{sec:flip}

In Figure~\ref{fig:multi}, we show the 12~epochs images
of 3C~84 at 43~GHz obtained with KaVA  
during the period from 2015 August to 2016 June.
 Here we can see the overall propagation of the C3 component 
 towards the south direction, and 
 the change of the flux densities of the C1 and C3 components.
\footnote{
 The apparent extended structure on the south-west side
 of the image in the epoch 2015 September in Figure~\ref{fig:multi}
 is not a real structure. It was an artifact by the poorer uv coverage
 because of the lack of  three stations in KaVA on this epoch
 (see the second column of Table~1).}
In Figure~\ref{fig:position}, we show the 
location of the peak position of the C3 component 
relative to the peak location of the C1 component.
The C3 position on August 1 2015 is well  along 
 the extended line of the past trajectory of C3 (Nagai et al. 2010; Suzuki et al. 2012; Hiura et al. 2018). 
As seen in Fig.~\ref{fig:position}, we discover 
the flip of the C3 position during August-September in 2015.
After September 2015, all the epochs obtained with KaVA and VLBA
show the flip of C3 in relative right ascension.
As seen in Figure~\ref{fig:multi},
the motion of the peak position of C3 was decelerated and C3 wobbled 
for a few months. 
After that, the peak position of C3 resume to propagate towards the south direction in the sky.
Intriguingly, the apparent velocity  of C3 temporally 
seems to reach the velocity comparable to $c$ after September 2015. 
Hence, doppler boosting could potentially contribute the brightening of C3 in the light curve. 
Since the number of the observation epochs here is not enough to conclude it, 
we will separately explore it  in our forthcoming paper
by adding further observation epochs in 2017-2018.

\subsection{Light curves of C1 and C3}

In Figure \ref{fig:flux},
we show the light curves of the C3 and C1 components. 
The clear monotonic increase of  the peak intensity of C3 
 is found in the overall trend of the light curve.
The peak intensity C3 increased  more than a factor of $\sim 4$.
On the other hand, the peak intensity of C1does not show
such a monotonic increase in the flux density and it was
in the range of $\sim 5-6$~Jy/beam during the observation period.
The peak intensity of C3 finally exceeded that of C1 during 2016 winter. 
In the last epoch, i.e.,  2016 June 12, the peak 
intensity of C3 became more than 4~Jy brighter than that of C1.
As shown in In Figure \ref{fig:flux} and Table~1, 
the peak flux density measured by VLBA on 
2015 Dec 5th and the one measured by KaVA on 2015 Dec 6th 
show a good agreement to each other.
This agreement of the independent VLBI arrays of  KaVA and VLBA 
well guarantees the robustness of the flux measurements of these two arrays.

It is intriguing to see that the onset of the C3 flux density increase
was seen about MJD~57350. It means that  the brightening of C3 
realized with $\sim 100$~days delay from the date of the flip around MJD~57250. 
As explained in \S~\ref{sec:flip}, the propagation of C3 suspended for a few months. 
After around MJD~57350, it restarted to advance to southward. 
The  onset of the C3 flux density increase is coincide with this restart timing 
of the C3 propagation towards the south direction rather than the moment of the flip of C3.


\section{Number density of the clump}

To change the direction of jet propagation,
there may be two possible mechanisms.
One is the  precession of jet axis, 
while the other is  the collision with obstacles.
As for the precession, 
 Falceta-Gon{\c c}alves et al. (2010)
proposed a scenario that the observed morphology of 3C 84 on 10-100 kpc-scales can be well explained
by a precessing jet with a period of $\sim 10^{7}$ years, using three-dimensional numerical simulations
considering the jet precession evolution.
Recently,an indication of the  precession of the 3C~84 jet 
has been reported by Hiura et al. (2018) 
based on VERA monitoring of 3C~84 at 22~GHz during the period of 2007-2013.
The VERA data revealed that the relative position of C3 against C1 showed  a smooth sinusoidal curve with
a period of about 6 years, which can be explained by the jet precession.
The relative position of C3 detected by KaVA during in 2015 reported in the present work, 
however, is clearly different from the sinusoidal curve but a flip and wobbling motion, 
which can be naturally 
explained by a collisiton between the jet and  surrounding non-uniform/clumpy medium.
The recent detection of  the enhanced  linearly-polarized emission 
of the C3 component in 3C~84 has been found in the same period 
and this enhancement can also support the  interaction between 
the jet and clumpy medium (Nagai et al. 2017, see also \S~5).
Because of these reasons, we assume that the flip of C3
is caused by the dense clump and hereafter we estimate the number density of the  clump.

If the 3C~84 jet thrust completely overwhelms the ram pressure by the clump,
then the clump is dynamically negligible. 
Then the jet would continue to propagate along the same direction  
without any flip of the jet head.
On the other hand, too dense clump should 
completely intervene  jet propagations.
To realize the flip of the jet head, 
the jet thrust would be comparable the ram pressure by the clump
(e.g.,  a numerical simulation shown in Figure~6 of the paper Wang et al. 2000).
From this condition, one can obtain
\begin{eqnarray}\label{eq:momentum-balance}
\frac{L_{\rm j}}{c}&\approx &\rho_{\rm cl} v_{\rm h}^{2}A_{\rm cross}  ,
\end{eqnarray}
where 
$L_{\rm j}$, 
$\rho_{\rm cl}$,  
$v_{\rm h}$, and
$A_{\rm cross}\equiv \pi (D_{L} \theta_{\rm flip})^{2}$ 
are
the total kinetic power of the jet, 
the mass density of the clump,
the jet-head advancing velocity and 
the effective cross sectional area of the jet-clump interaction, respectively
 (Begelman and Cioffi 1989; Kino and Kawakatu 2005).
 {In sub-Sec 5.2, we will present a consistency check about 
 this assumption of Eq. (\ref{eq:momentum-balance}).
According to recent VLBI observations, the averaged velocity of the jet 
after the emergence of the C3 component is known as
\begin{eqnarray}\label{eq:velocity}
0.2~c \lesssim v_{\rm h} \lesssim 0.3~c    ,
\end{eqnarray}
(Nagai et al.  2010; Suzuki et al. 2012; Hiura et al. 2018).
Here we set the effective cross sectional area as
 \begin{eqnarray}
A_{\rm cross}
\approx  1.9 \pi\times 10^{35} 
\left(\frac{\theta_{\rm flip}}{0.4~{\rm mas}}   
\right)^{2}~{\rm cm^{-2}}   .
\end{eqnarray}
In principle, the cross sectional area of 
the clump $A_{\rm cl}\equiv \pi R_{\rm cl}^{2}$
can be larger than this effective cross sectional one $A_{\rm cross}$
since there may exist a remaining part of the clump 
on the west side of the jet-head flip region.
Hence, we introduce 
a numerical factor $\eta\equiv   A_{\rm cl}/A_{\rm cross}$.
Therefore, we take into account of  such cases and  
hereafter we mainly examine a case of $\eta \approx 1$.
The corresponding radius of the clump ($R_{\rm cl}$) can be
given by $R_{\rm cl}\approx 0.14 \eta^{1/2} ~{\rm pc}$.
In the sub-section 5.2, we will briefly address the case of  $\eta \gg 1$ as well.

The total kinetic power of the jet $L_{\rm j}$ has been also constrained in the literatures.
 Heinz et al. (1998) argued that  the time-averaged total power of the jet in NGC~1275
 probably exceeds  $L_{j}\sim 1\times 10^{46}~{\rm erg~s^{-1}}$.
  They derived the conclusion based on the 
  observed properties of X-ray cavities in the central region
  of the Perseus cluster, which is supposed to be inflated by
  relativistic particles of the shocked jet.
  Hence, the minimum value of the jet power can be nominally
  $L_{j}\sim 1\times 10^{46}~{\rm erg~s^{-1}}$.
 \footnote
 { However,  they also suggested that the jet power in a 
  quiescent state (corresponding to off-state mentioned
  in the paper of Reynolds and Begelman 1997)
  may be lower than  $L_{j}\sim 10^{46}~{\rm erg~s^{-1}}$.}
As for the maximum value of  $L_{j}$,
Tavecchio et al. (2014) employed the largest jet power 
$L_{j}=2.3\times10^{47}~{\rm erg~s^{-1}}$
based on their model fitting against the  non-thermal emission spectrum
observed in 3C~84.
This value suggested by Tavecchio et al. (2014) 
is comparable $L_{\rm Edd}$. We adopt this value as the 
maximum value of   $L_{j}$.
Summing up, we set the range of  $L_{j}$ as 
\begin{eqnarray}\label{eq:power}
1\times 10^{46}~{\rm erg~s}
\lesssim  L_{\rm j} \lesssim
2\times 10^{47}~{\rm erg~s}
\end{eqnarray}
in the present work .


In Figure \ref{fig:density}, we show the estimated range of 
the clump density  $n_{\rm cl}$ 
based on Eq.~(\ref{eq:momentum-balance}), i.e.,  
simply based on the momentum balance relation.
By substituting the allowed values of $L_{\rm j}$ and  $\rho_{\rm cl}$
shown in Eqs.~(\ref{eq:velocity}) and (\ref{eq:power}), 
one can obtain the allowed value of $n_{\rm cl}$ and 
it is given by
\begin{eqnarray}\label{eq:n_cl}
4 \times 10^{3}~{\rm cm^{-3}}
\lesssim  n_{\rm cl} \lesssim
2 \times 10^{5}~{\rm cm^{-3}}   .
\end{eqnarray}
The bottom right part of the allowed region
in Figure \ref{fig:density} corresponds to the case of a denser clump, 
while the left side region represents the case of a thinner clump.
We add to note that
an enclosed total mass of the clump 
($M_{\rm cl, obs}=4\pi R_{\rm cl}^{3}/3$) can be estimated as
$M_{\rm cl, obs} \approx 3~M_{\odot} 
n_{\rm cl,4}(R_{\rm cl}/0.14~{\rm pc})^{3}$.

\section{Discussions}

\subsection{Origin of the clump}

In the previous section, we constrain 
the number density of the clump located at ~1~pc
from the central engine of 3C~84 via 
the simple dynamical argument of jet-clump interaction.
Here, we will discuss  possible origins of the clump based on its location and number density.
Since typical density and location of a cloud in a broad line region (BLR) 
(e.g., Osterbrock 1989;  Perterson 1997; Maiolino et al. 2010; Bentz et al. 2013; Czerny et al. 2017)
are obviously different from those for the clump in 3C~84, 
we will not further discuss the possibility of clouds in BLR below.

\subsubsection{A cloud in narrow line region?}

Let us discuss a possibility of a cloud in narrow line region (NLR) as the origin of the clump.
A typical  number density is known as  $\sim 10^{2-5}~{\rm cm^{-3}}$
(e.g., Osterbrock1989)
and thus it is obvious that a typical number density of 
cloud in NLR well agree with the estimated number density of the clump ($n_{\rm cl}$) in 3C~84. 
In this sub-section, we further discuss whether the redial dependence of 
$n_{\rm cl}$  in 3C~84 has a similar trend with those in other AGN sources.
Typically, NLR clouds  in  moderately luminous AGN such as 
Seyfert galaxies are expected to be  
located in the broad radial range of
$10^{1-3}$~pc distance from the central engine 
(e.g., Ramos Almeida and Ricci 2017).
Some of NLRs expand large enough to be partially resolved on the sky
(e.g., Pogge 1988; Tadhunter and Tsvetanov 1989; 
Schmitt et al.  1994; Fischer et al. 2013).

First, let us begin with an estimate the size of NLR ($R_{\rm NLR}$) for NGC~1275.
The $L_{H\beta}$ luminosity is estimated by
the luminosity of the  $H\alpha$ and $H\beta$ emission line luminosities
has the empirical relation of $3 L_{H\beta}\approx  L_{H\alpha}$
(Figure~5 in Greene and Ho 2005) and 
we have $L_{H\alpha}\sim 1\times 10^{42}~{\rm erg~s^{-1}}$
 in NGC~1275 by KANATA observation  (Yamazaki et al. 2013; Kino et al. 2016).
Therefore, the size of NRL in NGC~1275 can be estimated as
$R_{\rm NLR}\approx 0.6\times 10^{2}~{\rm pc}~
L_{H\beta,41}^{1/3}
\epsilon_{\rm fill, -2}^{-1/3}
n_{4}^{-2/3}$
where the filling factor of the NRL, and 
the number density of electrons are
$\epsilon_{\rm fill}$, and $n$, 
respectively.
The derived $R_{\rm NLR}\ $value is comparable to 
typical values in other sources
(e.g., Osterbrock 1989).


Next, let us compare the location and $n_{\rm cl}$ of the clump
in 3C~84 with those in other AGN sources.
Interestingly, Walsh et al. (2008)  conducted the
imaging spectrograph observations of nearby low-luminosity AGNs
with the Space Telescope Imaging Spectrograph (STIS) 
aboard the the Hubble Space Telescope (HST)
and  and they find 
 the electron number density measurements are characterized by a similar
slope with the normalization of the electron number density 
$5\times 10^{3}~{\rm cm^{-3}}$ at $\sim 1$~pc distance from the core
for five low-luminosity AGNs 
of NGC~1052, NGC~3227, NGC~3998, NGC~4278, and NGC~4579
(see Figure~7 in Walsh et al. 2008). 
Murayama and Taniguchi (1998) also indicate 
the radial dependence of the electron number density distribution in NLR
via detection of coronal lines, which is known as an indicator of 
denser clump.
Regarding the location of the clump, i.e., 1~pc distance from the central engine, 
the Figure~7 of Walsh et al. (2008)  indicates the electron number density
as $n_{\rm cl}\sim 10^{3-5}~{\rm cm^{-3}}$, which well agrees with our estimate.
Therefore we conclude that the clump in 3C~84 can be undersood as a cloud 
located deep inside (i.e., $\sim 1$~pc) 
the NLR with a normal number density.

\subsubsection{Self-gravitating molecular cloud?}

As alternative case which also agrees with the estimated $n_{\rm cl}$,
here we discuss the case when an intergalactic cloud
such as a dark cloud, a circum-stellar cloud, with its 
typical sizes, masses, and temperatures  
$0.1-3$~pc,
$10^{2-7}~{\rm cm^{-3}}$, and 
$10-100~{\rm K}$
(Goldsmith 1987)
is the origin of the clump. 
Interactions between galactic jet sources and such cold clumps are indeed
suggested in some observations (e.g., Fukui et al. 2009).
Here we briefly discuss  when the clump can be a self-gravitating one
since intergalactic molecular clouds are quite important in the context 
of  star formations.
The  radius and mass of a self-gravitating clump 
in the context of AGNs have been estimated in literatures
(e.g., Krolik and Begelman 1988; Honig and Berkert 2007, Kawaguchi and Mori 2011)
and we follow those arguments.
The two conditions required for self-gravitational clumps are as follows.
The first requirement is 
(1) the free-fall time for spherical clump
($t_{\rm ff}=\sqrt{3\pi/(32\rho_{\rm cl} G)}$) is shorter than 
the sound crossing time ($t_{\rm crs}=R_{\rm cl}/c_{s}$). 
The second requirement is 
(2) these clump
should be stable against the tidal force
in the gravitational field of the central black hole
($F_{\rm tidal} \approx 2GM_{\bullet} M_{\rm cl}R_{\rm cl}/r^{3}$)
by the self-gravity force ($F_{\rm sg} \approx GM_{\rm cl}^{2}/R_{\rm cl}^{2}$)
where $r$ is the distance from the central black hole to the clump.
The second requirement comes down to the relation
$(R_{\rm cl}/r)^{3} \lesssim M_{\rm cl}/2M_{\bullet}$.
From these conditions, we have 
$R_{\rm cl} \lesssim 
\frac{\pi}{\sqrt{16 G}}\frac{c_{s}r^{3/2}}{M_{\bullet}^{1/2}} 
\approx
1.6\times 10^{-2}
c_{s,5}
(r/1~{\rm pc})^{3/2} 
(M_{\bullet}/10^{8}~M_{\odot})^{-1/2} ~{\rm pc}$
and
$M_{\rm cl, crit} = \frac{\pi^{2} c_{s}^{2}}{8G}R_{\rm cl} \approx 
0.5~M_{\odot} 
c_{s,5}^{3}
(r/1~{\rm pc})^{3/2} 
(M_{\bullet}/10^{8}~M_{\odot})^{-1/2}$.
From this, it is clear that the clump can become self-gravitating one only when the clump is cold
(i.e., slow sound speed of order of $c_{s}\lesssim 10^{5}~{\rm cm/s}$),
although there are no direct evidence for cold clump in 3C~84 so far.
Therefore, observational explorations of cold clumps are essential to test this scenario.
Further discussions on cold gas will be made in the sub-section 5.3.2.

\subsubsection{A giant cloud?}

Here we briefly discuss a possibility of the existence of a clump
with its cross sectional area
$A_{\rm cl}$ larger than  that of $A_{\rm cross}$ (i.e., $\eta >1$).
If a giant molecular cloud with its size larger than a few parsec scale
(Goldsmith 1987) known in our Galaxy exists in NGC~1275,
then it may be one of the possible candidates for it.
A remaining part of such a large cloud may probably
hold back on the western side of  the radio lobe made of the 
shocked material escaped from C3 and produce a remnant structure
(see details in  Sec 5.2).
Do we find such a remnant feature? 
Here we speculate that the diffuse component C2 may be identical to the remnant of
the exhausted jet produced by the interaction of the jet and a large clump.
Although little attention has been paid to C2,
its existence was well known in literatures (e.g., Venturi et al. 1993; Dhawan et al. 1998)
and C2 does not show a systematic motion but shows random wobbling behavior.
Such behaviors could be explained by the collision with the large clump, 
which may cause a back-flows of the exhausted jet  
(Asada et al. 2006; Mizuta et al. 2010).
This scenario could be investigated through
high energy $\gamma$-ray emissions caused by strong shocks 
(e.g., Kino et al. 2011, 2017), 
although detailed investigations of such high energy $\gamma$-ray spectra 
is beyond the scope of this paper.

\subsection{Comparison with hydrodynamical simulations}

In this sub-section, we will make 
comparisons of the observational results with hydrodynamical simulations of 
jet propagations in surrounding ambient matter.

\subsubsection{Jet confinement condition in a uniform ambient matter}

First of all, it is important to check the availability of Eq.~(1)
based on  hydrodynamical simulations of jet propagations.
To this end, simulations of jet propagations in a uniform ambient matter
provide us a criterion for the availability of Eq.~(1).
A propagation of a jet in a uniform ambient matter is essentially governed
by the momentum balance between the jet thrust and the ram pressure from the ambient matter.
Hence, the physical quantity $L_{j}/n_{a}$ is the key  to understand  the jet propagations
where $n_{a}$ is the number density of the ambient matter.
It practically corresponds to the case of $\eta \gg 1$.
Hereafter, we discuss the comparison of 
the obtained $n_{\rm cl}$ in 3C~84 with hydrodynamical simulation
of De Young (1993) by focusing on the quantity $L_{j}/n_{a}$.
Let us begin with a brief summary of the hydrodynamical simulation of De Young (1993), 
which thoroughly examine the various combinations of $L_{j}$ and $n_{a}$
and investigated the condition of  jet-confinement by uniform ambient matter.
They examined the case of jet propagations with the total kinetic power
$L_{j} \approx 1\times 10^{45}$ erg/s and changing
$n_{a}$ and they found no jet confinement below $n_{a}\lesssim 3~{\rm cm^{-3}}$
 (Fig. 9 of De Young 1993).
They also tested the case of $L_{j}\approx 1\times 10^{44}~{\rm erg/s}$and 
 the jet is still "confined" when $n_{a} \gtrsim1.6~{\rm cm^{-3}}$  
 (Fig. 9 of De Young 1993).
Therefore, the condition for realizing  the jet confinement by the uniform ambient medium
based on the simulation of De Young (1993) can be written as
\begin{eqnarray}\label{eq:confinement}
\frac{L_{j}}{n_{a}}  \lesssim (6\times 10^{43} - 3\times 10^{44}) ~ {\rm erg~cm^{3}/s}.
\end{eqnarray}
The estimate of $n_{\rm cl}$ together with the adopted $L_{j}$ shown
in Eq.~(\ref{eq:power}) in 3C~84 in the present work
shows a good agreement with the result of De Young (1993), 
i.e., the estimated upper limit of $L_{j}/n_{\rm cl}$ 
is almost the same as  the criterion  value inferred from the numerical result of  De Young (1993).
Thus, the flip of the jet head of 3C~84 is well justified as a consequence in this jet clump collision.

It is worth to add that the condition of confinement shown in 
Eq.~(\ref{eq:confinement}) is almost consistent also with the previous work of 
Kawakatu et al. (2009), which shows that a hot spot, which corresponds to
the reverse shock at the tip of the jet,  
is generated when 
\begin{eqnarray}
\frac{L_{j}}{n_{a}}  \gtrsim 1\times 10^{44} ~ {\rm erg~cm^{3}/s}.
\end{eqnarray}
based on a Mach number value at the hot spot (termination shock) region.

\subsubsection{More realistic case: inhomogeneous ambient matter}

Next, we discuss the case of an inhomogeneous ambient matter, 
which would provide us with an insight for more realistic situation.
The work of Wagner et al. (2012) indeed made a  comprehensive work of jet propagation in 
two-phase interstellar medium (ISM), which consists of a warm and hot phases.
The warm phase has a density perturbation and it produces the clumps
with the upper limit temperature $\sim 10^{4}$~K.
This situation qualitatively corresponds to the case of $\eta \sim O(1)$.
 In Wagner et al. (2012), no lower temperature limit is enforced 
and temperatures  in the core of cloud (i.e. clump) may initially be less than 100~K.
They have found that the acceleration of the dense embedded clouds (clumps)
is realized by the ram pressure of the high-velocity flow through the porous channels of the warm phase.
This process transfers $10\%$ to $40\%$ of the jet kinetic energy to the cold and warm gas, 
accelerating/blowing it to velocities that match those observed radio jets in AGNs. 
That is so-call  AGN feedback in mechanical energy form.
The key feature seen in the simulation of 
is a mixture of the flip of the jet head and blown out of clumps  (Figure 9 of Wagner et al. 2012).
Because of the significant non-uniformity of ambient matter, dynamics 
of clumps cannot be purely 
determined by the head-on collision between the jet and a single clump alone.
Perhaps, such a hybrid case of the jet flip and AGN feedback may be more realistic in actual AGN jet sources.
Wagner et al. (2012) show that the AGN Feedback is efficient for
$ 10^{43}~{\rm erg~s^{-1}}\lesssim  L_{j} \lesssim10^{46}~{\rm erg~s^{-1}}$, 
with  the number density of the clump as 
 $30~{\rm cm^{-3}}\lesssim n_{\rm cl} \lesssim 10^{3}~{\rm cm^{-3}}$. 
Since there is an overlap in these parameter ranges, 
we can expect a hybrid phenomena in 3C~84.
Indeed, as mentioned in \S 3.1, we have detected a short-lived extended
structure at the West of C3 from 2015 December to 2016 January. 
This structure is similar with a substructure at the collision site between the jet and a clump 
demonstrated in the Figure 9 of Wagner et al. (2012) .
 As for the cloud/clump velocity, Wagner et al. (2012) predicted
 a typical velocity of the clumps after the collision as from a few $100$ km/s to 1000~km/s.
Therefore, this clump velocity will be one of the key observational quantities
towards testing AGN feedback process in 3C~84 in the future.

\subsection{Other signatures of clumpy medium}

Lastly, we briefly discuss other signatures of  
clumpy medium in 3C~84.

\subsubsection{Enhancement of linear polarization emission}

Nagai et al. (2017) recently reported the detection of 
a significant polarized emission at the C3 component by VLBA observation data
in the BU blazar monitoring program. 
Faraday rotation is also detected within an entire bandwidth of the 43~GHz band.
The obtained rotation measure is  $\sim 6\times 10^{5}~{\rm rad~m^{-2}}$ at a maximum. 
 Similar RM values were also reported at 210-345~GHz
 by Plambeck et al. (2014) using CARMA and SMA 
 while the 210-345~GHz emissions are likely to originate in the compact region at the close vicinity of the black hole.  
 Nagai et al. (2017) claimed that a simple spherical accretion flow 
 cannot explain the RM observed with the VLBA and SMA/CARMA consistently.  
To reconcile it, Nagai et al. (2017) proposed a local clumpy/inhomogeneous ambient medium
being responsible for the observed  RM. 
When an equipartition condition between magnetic field and accreting gas holds, 
the electron density is estimated as  $\sim 10^{4}~{\rm cm^{-3}}$, 
which is comparable to the estimate of $n_{\rm cl}$ in this work.
These results also strongly support  the existence of the clump near the C3 component.

\subsubsection{Searching for  cold gas}
Observationally,  direct detections 
of cold gas within ~10 pc scale from central BHs
are challenging and such explorations have just begun for other nearby active galaxies.
In this context, Imanishi et al.  (2018) made a first detection of cold gas on this 
scale (what they detected was a rotating dusty molecular torus) in radio band
by using the benefit of ALMA's high spatial resolution. 
First VLBI detection of HCN molecular absorption line features
 in NGC~1052 has been reported by Sawada-Satoh et al. (2016). 
We add to note that there are some detections of absorption lines by such clumps 
 for ultra-luminous IR galaxies (e.g., Geballe et al. 2006, Sajina et al. 2009).
 We further add to note that a large fraction of  X-ray selected Seyfert 2 galaxies
indeed show significant variations in the X-ray absorbing column density of $N_{\rm H}$
on the typical timescale less than 1~yr, which clearly 
suggests the presence of  clumpy circumnuclear material on a scale below a parsec
(Risaliti et al. 2002).

Although a cold gas has not yet detected at the circumnuclear region of NGC 1275, 
ALMA can potentially probe the cold molecular gas in the region.
We proposed ALMA observations to study the morphological
and kinematical properties of the cold molecular gas
in the central a few ten parsec (PI: H. Nagai), 
and the observations were partially completed. 
We also performs KVN observations to look for 
absorption line caused by the cold molecular gas
in the central a few ten parsec (PI: K. Wajima).
Such high resolution observations towards NGC~1275 will definitely accelerate our 
understanding on a cold gas at the center of nearby radio galaxies.

\section{Summary}

By performing the multi-epoch monitoring observation with KaVA at 43~GHz
in the period from 2015 August to 2016 June,
we explore the kinematics of the notable newborn bright  component C3 
located at the tip of the radio jet 3C~84. 
Here we summarize our main findings.

\begin{itemize}

\item
We discovered a flip of the  peak position of the C3 component during 2015 August-September.　
The amount of the positional shift in the angular scale was  about 0.4~mas, 
which corresponds to 0.14 parsec in physical scale.
After the flip, the C3 wobbled at the same location for a few months
and then it restarted to propagate towards the southern direction.
For consistency check purpose, we have analyzed three epochs of  
VLBA archival data of 3C~84 at 43~GHz and the obtained position of C3 
showed good agreement with those measured by KaVA.

\item

 We find the clear monotonic increase of the peak intensity 
of the C3 component during the observation period. The peak intensity of 
C3 increased more than a factor 4 at the end of the observation period. 
On the other hand, the peak intensity of C1 remained almost 
constant within the range of 5-6 Jy/beam during the period.
The flip of C3 together with this monotonic enhancement of its flux density
can be explained by  a collision between the jet and a dense clump.
The onset of the C3 flux density increase
was seen about MJD~57350, which is not coincide with the timing of the flip 
happened on around MJD~57250. 
The  onset of the C3 flux density increase was coincide with the restart timing 
of the C3 propagation to southward rather than the moment of the flip of C3.

\item
We estimate the density of the presumable clump based on the momentum balance
between the jet thrust and the ram pressure from the clump
and we find that the clump number density 
 is estimated about 
 $4\times 10^{3}~{\rm cm^{-3}}  \lesssim n_{\rm cl} \lesssim 2\times 10^{5}~{\rm cm^{-3}}$. 
The clump's location and estimated number density are in good agreement with
a cloud located deep inside the NLR or a dense portion of an intergalactic molecular cloud.

\item

The estimated value of  $L_{j}/n_{\rm cl}$ for the jet-clump collision in 3C~84 
satisfies the criterion for the jet confinement
derived by the hydrodynamical simulation done by De Young (1993).
It guarantees the availability  of Eq.~(1) and the jet flip by the
clump can be  naturally expected.
Further comparison with the hydrodynamical simulation
by Wagner et al. (2012) provides us predictions 
about more realistic case of the jet propagation in a inhomogeneous ambient gas.
Accelerations of clumps by the jet momentum can be
realized through the porous channels in the inhomogeneous ambient gas.
 Since there is an overlap in the parameter ranges of  Wagner et al. (2012), 
similar phenomena in 3C~84 may be expected.
 We will further perform observations exploring such phenomena  in the future.

\end{itemize}


\bigskip
\leftline{\bf \large Acknowledgment}
\medskip

\noindent

This work is mainly based on KaVA observations, 
which is operated by the the Korea Astronomy and Space Science
Institute (KASI) and the National Astronomical Observatory of Japan (NAOJ).
We thank K. Sugiyama for useful discussion.
This work is partially supported by  JSPS KAKENHI  
Grant Numbers JP18K03656 (MK) and JP18H03721 (KN, MK, KH).
NK acknowledges the financial support of MEXT KAKENHI (16K17670).
HN is supported by MEXT KAKENHI Grant Number 15K17619.
This study makes use of 43~GHz VLBA data from the VLBA-BU Blazar Monitoring Program (VLBA-BU-BLAZAR;
http://www.bu.edu/blazars/VLBAproject.html), funded by NASA through the Fermi Guest Investigator Program. 
VLBA is an instrument of the Long Baseline Observatory. The Long Baseline Observatory is a facility 
of the National Science Foundation operated by Associated Universities, Inc.
This work was supported by the National Research Council of Science \& Technology (NST) granted by the International joint research project (EU-16-001).


\appendix
\section{VLBA archival data analysis}

For consistency check,
we used the VLBA archival data taken as
a part of Boston University Blazar Program. 
The observations were done on 2015 August 
and 2015 September at 43~GHz with all ten VLBA stations. 
The data consists of 4 intermediate frequencies (Ifs) with 
a 64-MHz band width for each IF. Total bandwidth is 256~MHz per polarization. 
Both right-hand and left-hand circular polarizations were obtained.
The data calibration procedure is the same as the one 
performed against KaVA data. Hence, we do not repeat it here.
In order to compare the VLBA and KaVA images,
we restore the VLBA images with the gaussian beam size 
of 0.75~mas.
Overall structures and fluxes obtained in the two epochs are
well consistent with those obtained by KaVA.



\begin{figure} 
\includegraphics
[width=16cm]
{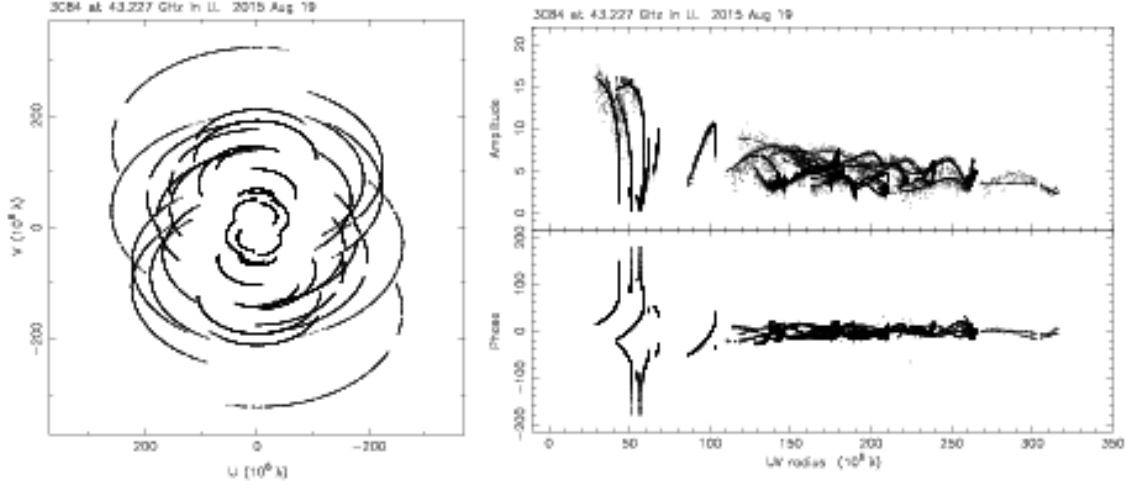}
\caption
{The $uv$ coverage and the visibility amplitude and phase 
obtained on 2015 August. This is one of the epochs
in this KaVA monitoring of 3C~84 at 43~GHz.}
\label{fig:visibility}
\end{figure}
\begin{figure} 
\includegraphics
[width=16cm]
{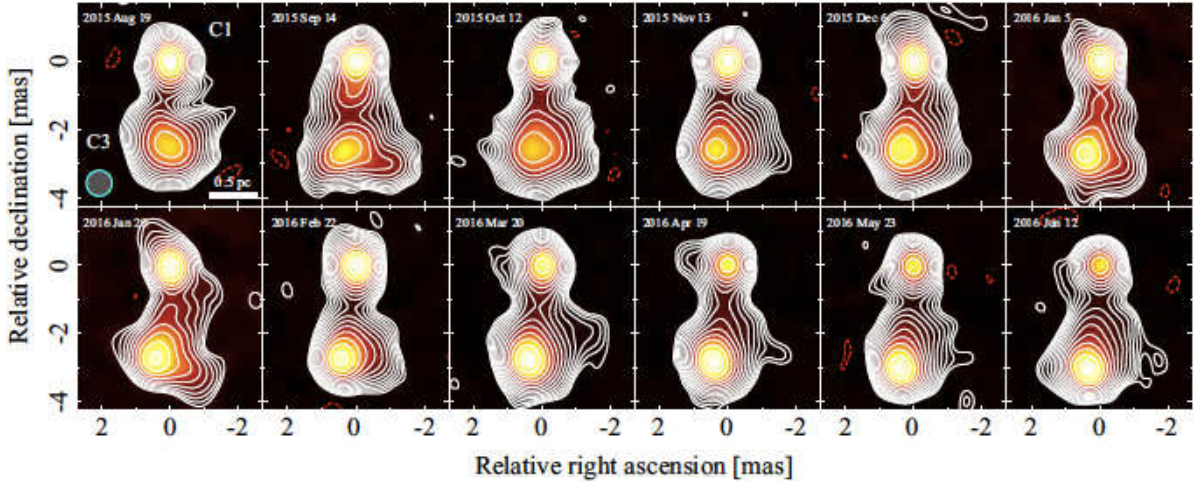}
\caption
{Multi-epoch KaVA images of 3C~84 at 43~GHz.
Twelve-epoch VLBI images of 3C~84 obtained by KaVA at 43.2~GHz from
2015 August 19 (top left) to 2016 June 12 (bottom right). All images
are restored with the circular Gaussian beam of 0.75~mas $\times$
0.75~mas, which is indicated in the lower left corner of the first-epoch
image. The lowest contour is three times the off-source rms noise
($\sigma$). The dashed and solid curves show negative and positive
contours, respectively, and the contour levels are $-3\sigma$,
$3\sigma \times (\sqrt(2))^n$ ($n$ = 0, 1, 2, ...).}
\label{fig:multi}
\end{figure}
\begin{figure} 
\includegraphics
[width=16cm]
{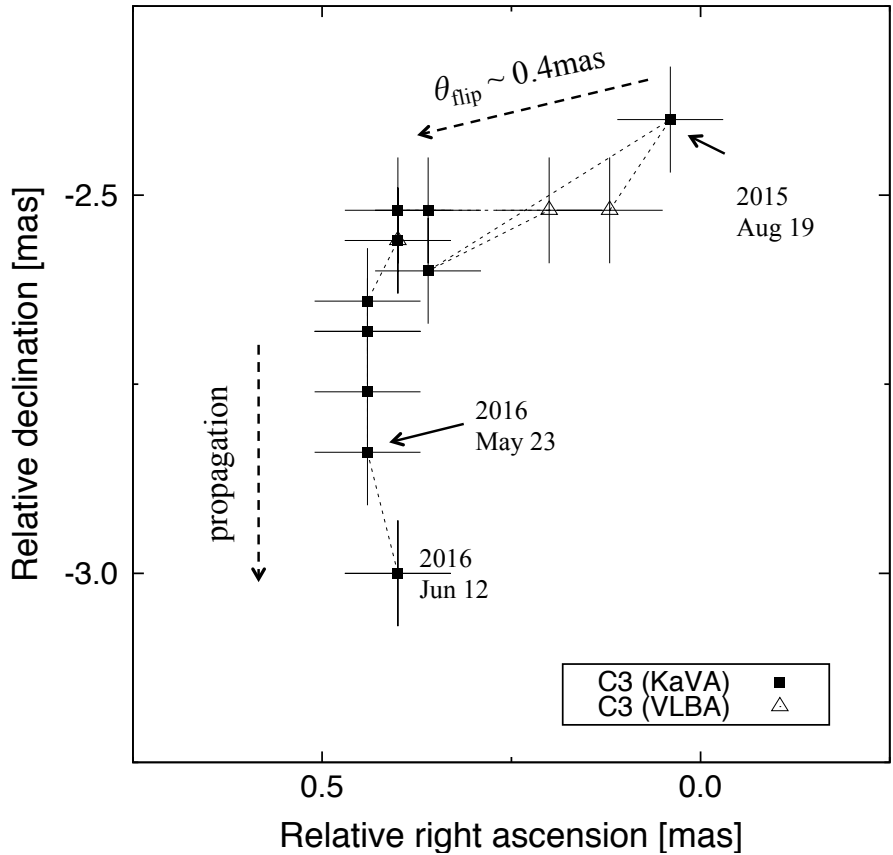}
\caption
{
Relative peak intensity position of the component C3 with respect
to C1 assuming as an origin (0, 0). The filled squares and open
triangles represent the peak intensity position obtained with KaVA
and VLBA observations, respectively.}
\label{fig:position}
\end{figure}
\begin{figure} 
\includegraphics
[width=16cm]
{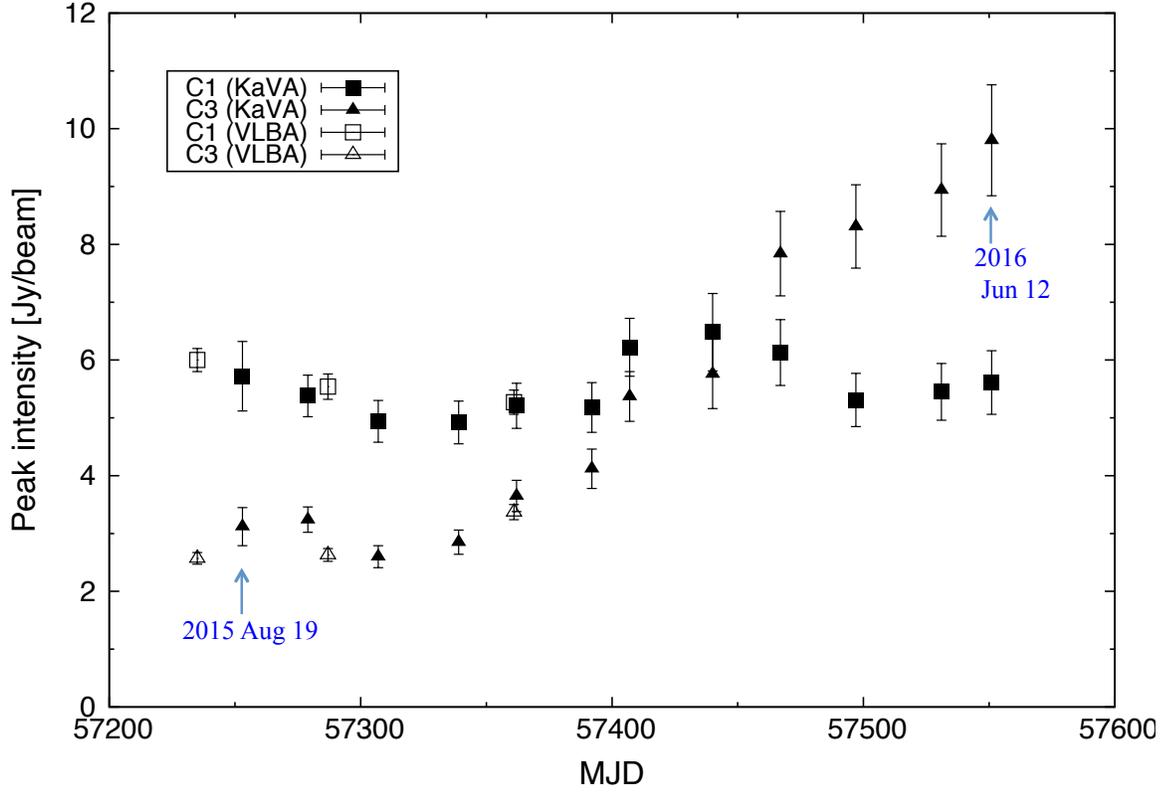}
\caption
{
Light curve of the components C1 and C3 measured by KaVA and VLBA
at 43~GHz from 2015 August to 2016 June. The filled and open points
correspond to the peak intensity of C1 (square) and C3 (triangle)
measured by KaVA and VLBA, respectively. All data points are obtained
from the restored images shown in Figures 2 and 6.}
\label{fig:flux}
\end{figure}
\begin{figure} 
\includegraphics
[width=16cm]
{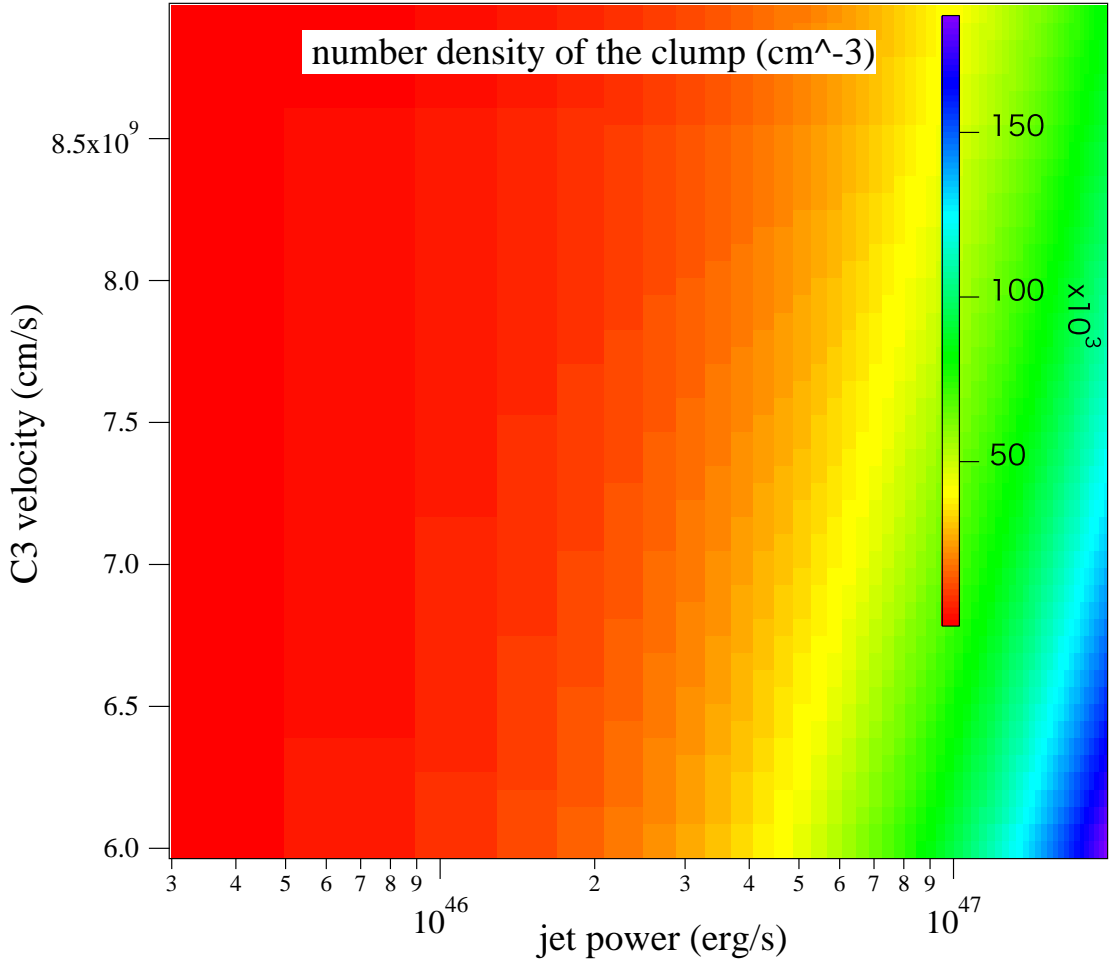}
\caption
{
The estimated number density of the clump ($n_{\rm cl}$)
based on Eq.~(\ref{eq:momentum-balance}).
The color contour indicates $n_{\rm cl}$ in the unit of $10^{3} \rm cm^{-3}$.
The vertical axis shows the advancing velocity of the C3 component
($v_{\rm h}$), while
the horizontal axis represents the total jet power ($L_{\rm j}$).}
\label{fig:density}
\end{figure}
\begin{figure} 
\includegraphics
[width=16cm]
{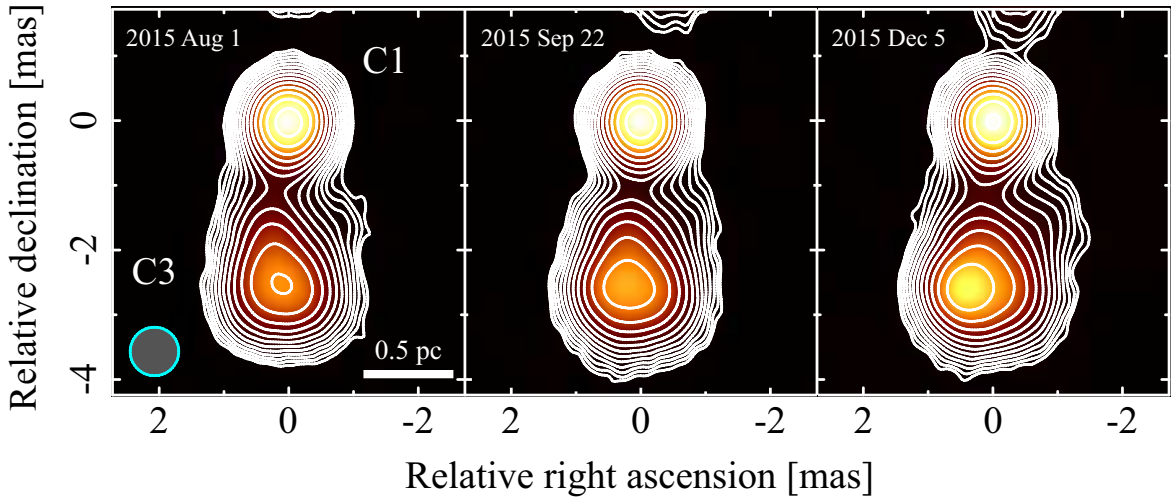}
\caption
{
Three-epoch VLBI images of 3C~84 obtained by VLBA at 43~GHz from
2015 August 1 (left) to 2015 December 5 (right). All images are
restored with the circular Gaussian beam of 0.75~mas $\times$ 0.75~mas,
which is indicated in the lower left corner of the first-epoch image.
Definition of the curves and contour levels are identical to that
shown in Figure 2.}
\label{fig:BU}
\end{figure}

\begin{table}
\centering
\caption{Observation summary}
\label{table:table1}       
\scalebox{1.0}{
\begin{tabular}{lccccc}
\hline\noalign{\smallskip}
UT date (MJD) & Stations & Image rms  & Peak intensity  & Dynamic range  \\
                         &                  & (mJy/beam)      &  (Jy/beam)  &        \\
\noalign{\smallskip}\hline\noalign{\smallskip}
  2015 Aug1 &
 VLBA (Boston U. monitor)&
0.90 & 
 6.00 &
$6.70\times 10^{3}$
 \\
 2015 Aug 19 (57253) &
 KaVA &
3.97&
5.72& 
$1.44\times 10^{3}$ 
 \\
  2015 Sep 14 (57279) &
 KaVA - MIZ, IRK, OGA &
4.95&
5.38& 
$1.09\times 10^{3}$
 \\
 2015 Sep 22 &
 VLBA (Boston U. monitor)&
 0.86 & 
 5.54 &
$6.44\times 10^{3}$ 
 \\
 2015 Oct 12 (57307) &
 KaVA - IRK &
 3.56 &
4.94 & 
$1.39\times 10^{3}$ 
 \\
  2015 Nov 13 (57339)&
 KaVA - MIZ &
 10.21&
4.92& 
$4.80\times 10^{2}$
 \\
 2015 Dec 5 &
VLBA (Boston U. monitor) &
0.97 &
5.27 &
$5.43\times 10^{3}$
 \\
 2015 Dec 06 (57362)&
 KaVA &
 2.28&
5.21& 
$2.29\times 10^{3}$
 \\
2016 Jan 05 (57392)&
 KaVA &
 3.49 &
5.18 & 
$1.48\times 10^{3}$ 
 \\
2016 Feb 22 (57440)&
 KaVA -KUS &
 4.93&
6.48& 
$1.31\times 10^{3}$
 \\
 2016 Mar 20 (57467)&
 KaVA &
 2.60&
7.84& 
$3.02\times 10^{3}$
 \\
2016 Apr 19 (57497)&
 KaVA &
 3.52&
8.31& 
$0.90\times 10^{3}$
 \\
2016 May 23 (57531)&
 KaVA-MIZ &
 9.95 &
8.94& 
$0.90\times 10^{3}$
 \\
 2016 Jun 12 (57551)&
 KaVA &
 6.67 &
9.80 & 
$1.47\times 10^{3}$ 
 \\
\noalign{\smallskip}\hline
\end{tabular}}
\end{table}


\end{document}